\documentstyle[aps,prb,multicol,graphicx,array]{revtex}

\begin{document}

\draft

\title{Thermal conductivity of Mg-doped CuGeO$_3$ at
very low temperatures:
Heat conduction by antiferromagnetic magnons}
%in the spin-Peierls phase}

\author{J. Takeya,$^{1}$ I. Tsukada,$^{1}$ Yoichi Ando,$^{1}$
T. Masuda,$^{2}$ and K. Uchinokura$^{2}$}
%\author{J. Takeya, I. Tsukada, and Yoichi Ando}
\address{${}^1$Central Research Institute of Electric Power
Industry, Komae, Tokyo 201-8511, Japan}

%\author{T. Masuda and K. Uchinokura}
\address{${}^2$Department of Advanced Materials Science,
The University of Tokyo,
7-3-1, Hongo, Bunkyo-ku, Tokyo 113-8656, Japan}

\date{Received \today}

\maketitle

\begin{abstract}
Thermal conductivity $\kappa$ is measured
at very low temperatures down to 0.28~K
for pure and Mg-doped CuGeO$_3$ single crystals.
The doped samples
carry larger amount of heat than the pure sample
at the lowest temperature.
This is because antiferromagnetic magnons
appear in the doped samples and are responsible for
the additional heat conductivity,
while $\kappa$ of the pure sample represents
phonon conductivity at such low temperatures.
The maximum energy of the magnon is estimated to be
much lower than the spin-Peierls-gap energy.
%which is consistent with theoretical proposal by Saito and Fukuyama
%[M. Saito and H. Fukuyama, J. Phys. Soc. Jpn. {\bf 66}, 3259, (1997)].
The result presents the first example that $\kappa$
at very low temperatures probes the magnon transport
in disorder-induced antiferromagnetic phase of spin-gap systems.

\end{abstract}


%66.70.+f Nonelectronic thermal conduction and heat-pulse propagation
%in solids
%75.30.Kz Magnetic phase boundaries
%75.50.Ee Antiferromagnetics
%
\begin{multicols}{2}
\narrowtext

Recently, impurity-substitution effect on
the spin-singlet ground state has been intensively studied
in a variety of low-dimensional spin systems,
and the results
indicate that only a slight substitution
of non-magnetic impurity essentially
changes the ground state.
When the disorder is introduced by the non-magnetic impurity,
antiferromagnetic (AF) ordering {\it immediately} appears
without destroying the spin-gap feature,
commonly in
a spin-Peierls (SP) compound CuGeO$_3$,
\cite{hase3,oseroff,regnault,sasago,manabe}
%,kojima}
a ladder compound SrCu$_2$O$_3$,\cite{azuma}
and a newly found
Haldane system PbNi$_2$V$_2$O$_8$.\cite{uchiyama,uchinokura}
Since
it was believed that
spin-singlet state and antiferromagnetism are mutually exclusive,
the case of CuGeO$_3$,
in which both
the remaining SP ordering and the AF
ordering developing with impurity-doping turned out to be
{\it long-range} order,\cite{regnault}
is remarkable.
Fukuyama {\it et al.} proposed
a theoretical model for this anomalous phase, i.e.,
dimerized antiferromagnetic (D-AF) phase.\cite{fukuyama1}
In their model,
staggered AF moments and lattice dimerization
are spatially modulated,
keeping the AF and SP long-range correlations;
degree of the lattice dimerization is larger in between the
impurity sites than around the impurity sites and
magnitude of the staggered moments are larger near the impurity sites.

\vspace{-3mm}

Even though the coexistence of the two long-range orders
are established in lightly-doped CuGeO$_3$,
it is to be elucidated how low-energy spin excitations of
SP state, which holds spin gap, and those of AF state,
which is gapless in isotropic systems, compromise with each other
in D-AF phase.
Recently, Saito and Fukuyama extended the theory of
Ref.~\onlinecite{fukuyama1} and predicted that a \lq\lq slow"
gapless AF-magnon branch shows up in addition to the gapped SP mode
when impurity is doped.\cite{saito1}
They predicted that the energy scale of the magnon branch is
smaller than the SP gap ($\sim$~24~K).
%contrastingly to magnons in usual uniform N\'{e}el state,
%whose energy can reach as high as the order of intrachain
%interaction energy $J_c$ ($\sim$~120~K \cite{nishi}).
Existence of this in-gap magnon branch is observed
by neutron scattering for rather highly substituted samples.
\cite{martin,hirota,gehring}
(Zn concentration was 3.2\% for the sample used in
Ref.~\onlinecite{martin}, for example).
However, since
recent study using high-quality Mg-doped single crystals,
in which impurities distribute more homogeneously than
in Zn-doped systems,
has revealed
disorder-induced first-order transition
around the Mg concentration of $x_c = 2.4\%$,\cite{masuda}
only below which the long-range SP ordering is established,
\cite{masuda2}
it is essential to examine the magnetic excitations
of samples
with impurity concentration less than $x_c$.
In this work, we have measured
thermal conductivity $\kappa$ at very low temperatures,
using lightly Mg-doped samples.
The samples are cooled
down to $^3$He temperatures,
in order to study the low-lying excitations.
The thermal conductivity of the doped
samples exceeds that of the pure sample at the lowest temperature.
We will show that
the {\it in-gap} magnons, which is intrinsic to D-AF phase,
indeed exist and are responsible
for the excess low-temperature heat
transport in the doped samples.

%\vspace{-30mm}

%We use Cu$_{1-x}$Mg$_x$GeO$_3$ single crystals,
%in which impurities distribute homogeniously
%compared to the Zn-doped system,\cite{masuda2}
%for the $\kappa$ measurement.
%which is an advantage in the establishment of the D-AF long-range
%order.\cite{masuda,masuda2}
The Cu$_{1-x}$Mg$_x$GeO$_3$ single crystals
were grown with a floating-zone method.
The Mg-concentration is determined by
inductively coupled plasma-atomic emission spectroscopy
(ICP-AES).\cite{masuda,masuda2}
For the thermal conductivity
measurement,
we use pure, $x = 0.016$, and $x = 0.0216$ samples,
all of which were already well characterized
using dc susceptibility
and synchrotron x-ray diffraction measurements.
\cite{masuda,masuda2,wang}
The transition temperatures are shown in Table~\ref{table1}
for each sample.
SP long-range order is observed
at $T_{\rm SP}^{\rm LRO}$
as a resolution limited FWHM of the x-ray Bragg peak
from lattice dimerization.
The N\'{e}el temperature $T_N$ is determined by the magnetic
susceptibility.\cite{masuda2}

%\vspace{-2mm}

Thermal conductivity is measured
down to 0.28~K with
$^3$He refrigerator using \lq\lq one heater, two thermometers"
technique.
Gold wires, which are tightly connected with
\begin{table}[bth]
\caption{
%Characteristics parameters of the measured samples.
\label{table1}}
\begin{tabular}{cccccc}
\tableline
\tableline
Sample
&$x$&$T_{\rm SP}^{\rm LRO}$~[K]&$T_N$~[K]&$\bar{w}$~[mm]&
$v_m^c$~[m/s]
%&$\frac{\pi v_m^c}{4 c}$~[meV]
\\
\tableline
A&0&14.5& &0.17\\
B&0.016&10.5&2.5&0.17&70
%&0.12
\\
C&0.016&10.5&2.5&0.24&70
%&0.12
\\
D&0.0216&8.5&3&0.10&140
%&0.23
\\
\tableline
\tableline
\end{tabular}
\end{table}
\begin{figure}[tbh]
\includegraphics[width=0.9 \columnwidth]{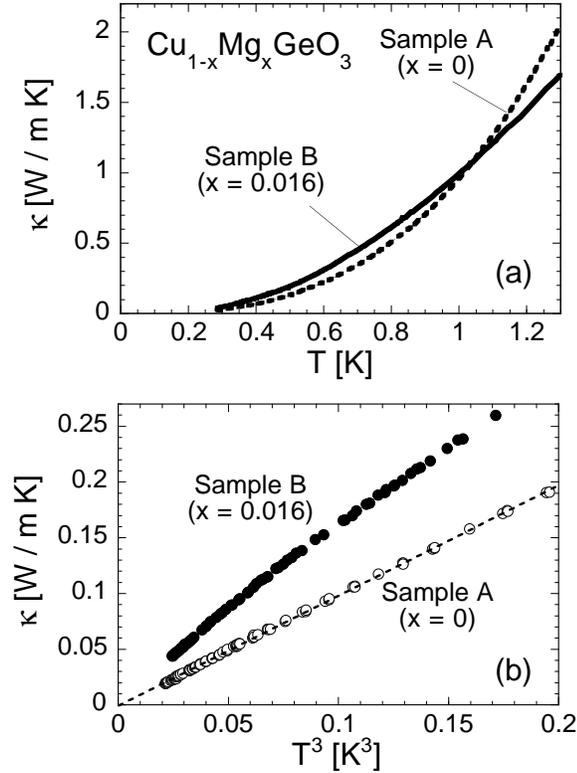}
\caption{(a) Thermal conductivity of
Cu$_{1-x}$Mg$_x$GeO$_3$ single crystals below 1.3~K.
Solid curve is for one of the $x = 0.016$ samples (B)
and dashed curve is for the pure sample (A).
(b) Thermal conductivity of the
same sample as a function of $T^3$.
The dashed lines show that $\kappa$ is proportional
to $T^3$ below 0.58~K for Sample A.
%and below 0.46~K for Sample B.
}
\label{fig1}
\end{figure}
\noindent
a microchip heater
and two calibrated RuO sensors,
are
attached on the samples by GE-varnish.
%The base of the crystal is anchored to a copper
%block.
Temperature difference between the two thermometers
are typically 3\% of the sample temperature.
Since we will discuss the thermal conductivity
in the low-temperature limit (Casimir's limit),\cite{casimir,berman}
where heat carriers are scattered dominantly by crystalline boundaries,
we paid special attention both to the sample size
and to the smoothness of the boundaries.
%Typical size of the crystals is $0.03 \times 1 \times 3$~mm$^3$
%($a \times b \times c$) and
Typical dimension along the $c$-axis is around 3~mm
and thermal gradient is applied along the $c$ axis.
As shown in Table~\ref{table1}, the geometrical mean widths $\bar{w}$
(square root of the cross section),
which is proportional to the mean free path in the Casimir's limit,
\cite{casimir}
of the pure sample (Sample A) and one of the $x = 0.016$ samples
(Sample B)
are set identical for direct comparison in $\kappa$.
%Note that
%it is essential to use
%samples with small enough $\bar{w}$
%in order to reach the Casimir's limit in the measured temperature
%region.
%We have prepared the samples with $\bar{w}$ of typically 0.2~mm.
The smooth boundaries of the crystal is achieved
by cleaving for the {\it b-c} surfaces and
by cutting with a sharp razor blade for the {\it a-c} surfaces.
Also, we have measured specific heat $C$,
in order to derive the
mobility of the relevant heat carriers.
The specific heat measurement is carried out down to 0.4~K
with the commercial
PPMS (Quantum Design) heat-capacity probe,
using a relaxation method.
The mass of the samples is typically 3~mg.
The samples used for the specific heat and the thermal
conductivity measurements are cut from the same piece of the crystals.

Figure 1(a) shows temperature dependence of $\kappa$ for
Samples A and B below 1.3~K.
In Fig. 1(b), $\kappa$ in the temperature range below 0.58~K
($T^3 < 0.2$~K$^3$) is plotted as a function of $T^3$.
$\kappa$ of the pure sample rapidly decreases with
decreasing temperature and becomes proportional to $T^3$
below 0.58~K.
At 1.3~K, $\kappa$ of the $x = 0.016$ sample is
smaller than that of the pure sample, so that our
previous results in higher temperature range are reproduced.
\cite{takeya}
It is natural that $\kappa$ is suppressed
in the presence of impurities owing to the scattering
by disorders.
However,
$\kappa$ of the $x = 0.016$ sample exceeds
that of the pure sample below 1~K down to the lowest
temperature,
which is not explained by the above simple picture.

First, let us discuss $\kappa$ of the pure sample.
Magnetic excitations are negligible in the pure sample
below 1.3~K,
because the spin-gap energy of CuGeO$_3$ is more than one
order of magnitude larger than the temperatures.\cite{incoherent}
Therefore, the heat is dominantly carried by phonons there.
Assuming kinetic approximation,
the phonon thermal conductivity $\kappa_{\rm ph}$ is written as
\begin{equation}
\kappa_{\rm ph} = \frac{1}{3} C_{\rm ph} v_{\rm ph} l_{\rm ph},
\end{equation}
where $C_{\rm ph}$ is specific heat,
$v_{\rm ph}$ is velocity and $l_{\rm ph}$ is mean free path
of the phonons.
Since $C_{\rm ph}$ also depends on temperature as $T^3$
at such low temperatures,
$l_{\rm ph}$ should be independent of temperature below 0.58~K.
This result means that the phonon conductivity reaches
the Casimir limit,
where
the mean free path is determined simply by the dimension of the crystal.
\cite{casimir,berman,defect-scat}
For a rectangular-shaped crystal
$l_{\rm ph}$ is given as
\begin{equation}
l_{\rm ph} = 1.12 \bar{w},
\end{equation}
%
%where $\bar{w}$ is the geometrical mean width of the crystal,
assumimg isotropic phonons.
\cite{thacher}
We have measured the specific heat independently
and examined the validity of Eqs. (1) and (2).
First, $v_{\rm ph}$ is calculated from $\kappa$ and $C$ data
by Eqs. (1) and (2), as
$v _{\rm ph} = (3 \kappa ) / ( C \cdot 1.12 \bar{w} ) \sim$~1600~m/s.
\cite{soundv}
On the other hand, $v _{\rm ph}$ can be estimated only from
the low-temperature $C$ data, assuming the Debye model.
Thus obtained value of $v _{\rm ph}$ is $\sim$~1800~m/s,
which is in good agreement with the estimation
from the $\kappa$ data.
The result shows that the relation of Eqs. (1) and (2)
is satisfied at the temperatures shown in Fig.~1(b) for
the crystal used.

Noting that phonon conductivity governed by
the boundary scattering gives
the maximum value of $\kappa_{\rm ph}$
(any additional scattering would suppress the conductivity),
we can notice that $\kappa_{\rm ph}$ of Sample B
cannot exceed $\kappa$ of Sample A in Fig. 1(b),
using Eqs. (1) and (2),
because $\bar w$ of Sample A and B
is identical (Table~\ref{table1}) and
little $x$ dependence is expected for $C_{\rm ph}$
and $v_{\rm ph}$.
%We can estimate $\kappa_{\rm ph}$ for the Mg-doped samples
%from $\kappa$ data of the pure sample
%using Eqs. (1) and (2)
%(as long as the Casimir's condition is satisfied),
%because little $x$ dependence is expected for $C_{\rm ph}$
%and $v_{\rm ph}$.
%Since $\bar w$ of Sample A and B
%is identical (Table~\ref{table1}),
%$\kappa$ of Sample A in Fig. 1(b) corresponds to
%$\kappa_{\rm ph}$ of Sample B.
Therefore, the result of the excess $\kappa$ in Sample B
requires an additional excitations
which can carry heat at low temperatures down to 0.28~K
in the Mg-doped sample.
Considering the AF ordering in the impurity-doped CuGeO$_3$,
it is most likely that antiferromagnetic magnons
are responsible for this excess low-temperature
heat conductivity in the Mg-doped sample.

\begin{figure}[tbh]
\includegraphics[width=0.9 \columnwidth]{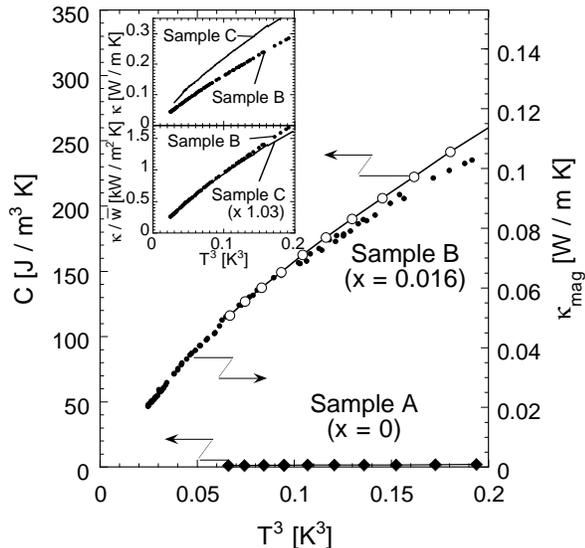}
\caption{Specific heat and thermal conductivity of
Cu$_{1-x}$Mg$_x$GeO$_3$ with the concentration of $x = 0.016$
below $\sim 0.58$~K as a function of $T^3$.
Specific heat of the pure CuGeO$_3$ is plotted together.
Inset: $\kappa / \bar{w}$ of two $x = 0.016$ samples
with different $\bar{w}$, i.e., Samples B and C.
}
\label{fig2}
\end{figure}

In order to examine whether the Mg-doped sample also
satisfies the Casimir's condition,
under which more quantitative discussion is possible,
we compare low-temperature $\kappa$ of Sample B and that of
another $x = 0.016$ sample with different $\bar{w}$ (Sample C).
The upper inset of Fig. 2 shows that the values of $\kappa$
are different between Samples B and C.
On the other hand, as shown in the lower inset,
the low-temperature $\kappa / \bar{w}$ data
($T^3 < 0.1$~K) of the two samples show only 3\% difference,
which is mainly due to the error in determining the distance between
the gold wires connected to the thermometers.
Since the result means that $\kappa$ differs in proportion to
$\bar{w}$ at low temperatures,
it is strongly suggested that the Casimir's condition is
satisfied for both samples and that both $\kappa_{\rm ph}$ and
the magnon heat transport $\kappa_m$
are governed by the boundary scattering there, i.e.,
\begin{equation}
\kappa_m = \frac{1}{3} C_m v_m^c 1.12 \alpha_c \bar{w}
\end{equation}
[$\alpha_c$
($\equiv \sqrt[3]{v_m^a v_m^b/{v_m^c}^2} \sim
\sqrt[3]{|J_a J_b|/J_c^2}$)
is a factor due to the anisotropy of the magnon velocity \cite{berman}
and $v_m^i$  is the magnon velocity in the $i$-th-axis direction).

%This idea is supported by the $\kappa$ data
%in the low temperature limit as follows.
%Note that $\kappa$ of the Mg-doped sample,
%which is considerably larger than that of the
%pure sample,
%also shows $T^3$ behavior below 0.37~K ($T^3 < 0.05$~K$^3$),
%meaning the additional thermal conductivity
%depends on temperature as $\propto T^3$
%in the low-temperature limit.
%Since specific heat of three-dimensional AF-magnons $C_m$
%is proportional to $T^3$,
%it is most likely that the excess heat
%is carried by 3-D AF-magnons and that
%the magnon transport $\kappa_m$ below 0.37~K
%is in the Casimir's limit,
%where the similar relation to $\kappa_{\rm ph}$ is satisfied, i.e.,

Assuming Eq. (3),
we can crudely estimate $v_m^c$
with the value of $C_m$ obtained by the independent
specific-heat measurement and examine
whether the magnon branch is {\it in} the SP gap.
The specific heat of Samples A and B is shown in Fig. 2
as a function of $T^3$ in the temperature range
below $\sim$~0.58~K$^3$ ($T^3 < 0.2$~K).
Since $C$ of the pure sample (diamonds in Fig. 2),
which represents $C_{\rm ph}$ in this temperature region, is
more than one order of magnitude smaller than that of
the Mg-doped samples, we can neglect $C_{\rm ph}$ and
assume $C_m \sim C$ for the doped sample.
$\kappa_m$ of the $x = 0.016$ sample is estimated
by subtracting $\kappa$ of the pure sample
and is plotted together in Fig.~2.
The $\kappa_m(T)$ data show exactly the same temperature
dependence as the $C(T)$ data below 0.46~K ($T^3 < 0.1$~K),
indicating again that
Eq.~(3) is satisfied in this temperature range.
Assuming $J_c \sim 10 J_b \sim 100 |J_a|$ and
$\alpha_c \sim 0.1$,\cite{nishi}
$v_m^c$ is estimated approximately
to be 70~m/s from Eq.~(3).
The AF-magnon energy is lower
than $q_c v_m^c$ $\sim 0.12$~meV
[$q_c$ ($= \pi / 4 c$) is the wave number at the center
of the Brillouin zone
and $c$ ($= 2.9$~\AA) is the distance between adjacent Cu
atoms along the Cu-O chain].
Since this value is two orders of magnitude
lower than the SP gap ($\Delta_{\rm SP} \sim 2$~meV),
we can conclude that the additional heat transport observed
in the Mg-doped CuGeO$_3$ is due to the in-gap AF-magnons.
\cite{delta_aniso}

\begin{figure}[tbh]
\includegraphics[width=0.9 \columnwidth]{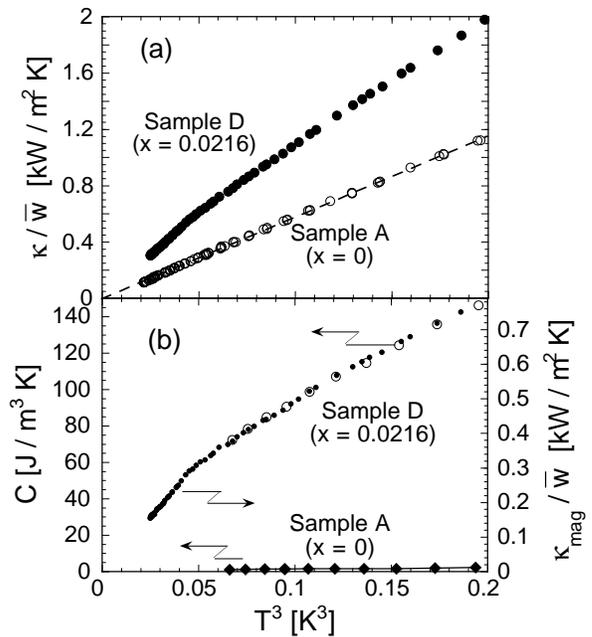}
\caption{(a) Thermal conductivity
of Cu$_{1-x}$Mg$_x$GeO$_3$ ($x = 0$, 0.0216)
crystals,
divided by $\bar{w}$
is plotted against $T^3$.
%The dashed line for Sample D indicate that
%$\kappa$ becomes proportional
%to $T^3$ at the lowest temperature ($T^3 < 0.05$~K$^3$).
(b) Specific heat and thermal conductivity of
Cu$_{1-x}$Mg$_x$GeO$_3$ ($x = 0.0216$)
below $\sim 0.2$~K as a function of $T^3$.
Specific heat of the pure CuGeO$_3$ is plotted together.
}
\label{fig3}
\end{figure}

The magnon velocity can be estimated also
for the $x = 0.0216$ sample (Sample D) in the same manner.
In order to subtract $\kappa_{\rm ph}$,
it is convenient to compare
$\kappa / \bar{w}$ of Sample D
to that of Sample A, which represents the
phonon contribution.
$\kappa / \bar{w}$ of Samples A and D is plotted against $T^3$
in Fig. 3(a).
$\kappa / \bar{w}$ of Sample D
is larger than that of the pure sample.
The difference corresponds to
the magnon contribution $\kappa_m / \bar{w}$
of Sample D.
In Fig. 3(b),
$\kappa_m / \bar{w}$ and independently obtained $C$ data
are plotted together.
Following the same discussion as that for the
$x = 0.016$ sample,
$v_m^c$ is estimated approximately
to be 140~m/s.
The upper limit of the magnon energy $q_c v_m^c$ is
$\sim 0.23$~meV, which is
twice as high as that of the $x = 0.016$
sample but is still much
lower than the $\Delta_{\rm SP}$.

%\vspace{-0.6mm}

The velocity of AF-magnon for the usual uniform
N\'{e}el state is given by $2 z J_c S c$,\cite{kittel}
where $z$ is a factor of order unity, $J_c$ is the interaction
energy and $S$ is the spin in each magnetic site.
Even when quantum fluctuations, which diminish
the effective value of $S$ to 0.2 times smaller,\cite{hase4}
is taken into account, the magnon velocity of
as fast as $\sim$~1000~m/s is expected,
(assuming $J_c = 120$~K).\cite{nishi}
Since we obtained much smaller $v_m^c$,
it is shown that effective $S$ is significantly
suppressed owing to the SP ordering.
In the model of Refs.~\onlinecite{fukuyama1} and \onlinecite{saito1},
the staggered moment is strongly suppressed
in between the impurity sites.
As the result, a sort of spatially \lq\lq averaged" spin,
which is directly proportional to the magnon velocity,
becomes much smaller than 1/2.
%In Ref.~\onlinecite{saito1}, the values of $v_m^c$ were calculated
%to be approximately
%90~m/s and 210~m/s for the $x = 0.016$ and $x = 0.0216$
%impurity-concentrations, respectively.
Our crude estimation from the low-temperature
$\kappa$ gives $v_m^c$ of the $x = 0.0216$ sample approximately twice
as large as that of the $x = 0.016$ sample,
indicating that $v_m^c$ rapidly increases with $x$.
Such $x$ dependence is consistent with the calculation
in Ref.~\onlinecite{saito1}.
Note that $v_m^c$ of around 1300 m/s
can be estimated for the Zn-3.2\%-doped sample from the neutron data,
\cite{martin}
which is nearly one order of magnitude larger
than the value estimated for our $x = 0.0216$ sample.
%and the values themselves are rather close.
%The results provide a strong evidence to
%the existence of the slow magnons, which is intrinsic
%to the coexistence of SP and AF long-range orders.

It should be emphasized that
the mean free path of the AF-magnon in the SP gap
reaches as long as
$1.12 \alpha_c \bar{w} \sim$~18~$\mu$m, according to Eq.~(3), i.e.,
the magnons are mobile to a distance 60,000 times longer
than the spin-distance ($c$) without being scattered.
Such coherent motion of the magnons is only possible with
extremely long correlation length of the corresponding
magnetic order.\cite{halperin}
Therefore,
the mixture of SP and AF ordering in the doped CuGeO$_3$
is certainly a true long-range order.

In summary, we have measured thermal conductivity
of pure and Mg-doped CuGeO$_3$ single crystals at very
low temperatures,
% down to 0.28~K,
in order to examine the anomalous low-temperature phase
in the impurity-doped CuGeO$_3$, where SP and AF order
coexists.
While the low-temperature thermal conductivity is dominated
by $\kappa_{\rm ph}$ in the pure CuGeO$_3$,
magnons also carry considerable amount of heat
in the Mg-doped samples.
Estimating the magnon velocity, we have shown that the
AF-magnons are present {\it in} the SP gap,
as predicted by Saito and Fukuyama.
It is demonstrated that thermal conductivity is
a powerful tool in elucidating
the low-energy magnetic excitations
in the disorder-induced AF phase of spin-gap systems.
Our next step is
to examine the excitations in the uniform AF phase
in highly substituted CuGeO$_3$,
and to seek differences from that of D-AF phase.

We acknowledge H. Fukuyama, who first suggested this experiment.
We also thank M. Saito, M. Nishi and M. Ishikawa
for helpful discussions.
The work done at the University of Tokyo is supported in part
by the Grant-in-Aid for COE Research of the Ministry of
Education, Science, Sports, and Culture.
%
%
% Place here the list of the references:
%

\end{multicols}


\begin{references}
\bibitem{incoherent}
In Ref.~\onlinecite{manabe}, considerable magnetic susceptibility
is observed below $\sim$~1~K,
because of incoherent magnetic excitations due to
residual impurities and/or defects.
However, such incoherent excitations do not
contribute to heat transport.
Also, magnetic specific heat is negligibly small, as
will be shown later (Fig. 2).
\bibitem{defect-scat}
Defect scattering would give
$T$-dependence of the mean free path.
Since scattering rate $\Gamma$ of the defect scattering
decreases with temperature,
it will exceed $\Gamma$ of the boundary scattering
at the low-$T$ limit, so that
the Casimir's condition is satisfied.
See Ref.~\onlinecite{berman}, p. 73, for example.
\bibitem{delta_aniso}
Although an anisotropy gap $\Delta_{\rm an}$ of around 1~K is
reported for a Zn-3\%-doped sample,\cite{martin}
the large magnetic specific heat observed at least down to $\sim$~0.4~K
(Fig.~2)
indicates that $\Delta_{\rm an}$ is suppressed in the lightly doped
samples to a value
comparable to or smaller than the lowest temperature of our $\kappa$
measurement.
It is reported for Si-doped samples that $\Delta_{\rm an}$
increases with $x$.\cite{hirota,gehring}
\end{references}
\end{document}